\documentclass[subeqnarray]{elsart}
\usepackage{graphicx}
\usepackage{natbib}
\usepackage{subeqnarray}

\usepackage{amssymb}

\begin{document}

\begin{frontmatter}


\title{Modeling of Negative Autoregulated \\
Genetic Networks in Single Cells}
\author{Azi Lipshtat, Hagai B. Perets, Nathalie Q. Balaban and Ofer Biham}
\address{Racah Inst. of Physics, The Hebrew University, 
Jerusalem 91904, Israel}




\begin{abstract}
We discuss recent developments in the modeling of negative
autoregulated genetic networks. In particular, we consider
the temporal evolution of the population of mRNA and proteins
in simple networks using rate equations. In the limit of low copy 
numbers fluctuation effects become significant and more 
adequate modeling is then achieved using the master equation formalism.
The analogy between regulatory gene networks and chemical reaction
networks on dust grains in the interstellar medium is discussed.
The analysis and simulation of complex reaction networks are also considered.
\end{abstract}

\begin{keyword}
genetic networks \sep repression \sep master equation
\end{keyword}
\end{frontmatter}

\section{Introduction}
\label{Sec:intro}

Recent advances in molecular biology techniques for the engineering of synthetic
networks have made possible the measurement of populations of mRNA's
and proteins in simple genetic networks.
Measurements of the average protein content of cells 
and their time dependence enabled to quantify the behavior of genetic
networks
\citep{Kalir2001}. 
These measurements have been modeled using rate equations,
mainly under quasi steady state conditions.
However, real biological systems are likely be away from steady state
\citep{Smith1968,Murray1989}.
Furthermore, many components of cells appear in low copy numbers and
are therefore subjected to large fluctuations.
Recently, such fluctuations at the level of a single cell
were measured experimentally using the  
green fluorescent protein (GFP)
\citep{Elowitz2002,Swain2002,Paulsson2004}. 
Measurements of protein levels in single cells revealed distributions
that depend on the topology of the regulatory 
network controlling the particular protein. 
For example, it was shown that negative autoregulated networks
reduce fluctuations
\citep{Becskei2000}. 
The modeling of these fluctuations cannot be done using rate
equations and requires the master equation formalism
\citep{McAdams1997,McAdams1999,Paulsson2000,Paulsson2000b,Kepler2001,Paulsson2002,Paulsson2004}. 
 
In this paper we consider the modeling of negative
autoregulated genetic networks in cell populations and in single cells. 
We focus on the simplest network in which
a single protein serves as a repressor for the production of its own mRNA.
Such network may serve as a module 
or ``network motif''
in complex regulatory networks
\citep{Milo2002,Milo2004}. 
We describe the time dependence of the system using rate equations. 
In commonly used models it is assumed that the population 
of the bound repressor proteins is in quasi steady state. 
We consider the dynamics of the network when this assumption does 
not hold. 
We show that in such cases the commonly used models underestimate
the response of the system to variations in the external conditions.   
In such cases one should take into account the bound repressors as
a separate population.
In the limit of low copy numbers of the mRNA's and proteins stochastic
noise becomes significant. We show that in this limit the rate equations
should be replaced by a master equation.
The rate and master equations used in the analysis of genetic networks
are closely related to those that describe chemical reaction 
networks on small grains.
In this context, the limit of low copy number is achieved for reaction
networks on interstellar dust grains, due to the sub-micron size
of the grains and the extremely low flux due to the low density
of the interstellar gas.
This analogy is discussed and results obtained for grain chemistry,
which may also be useful for genetic network analysis,
are presented.

The paper is organized as follows: in Sec. 2 we consider the dynamic behavior 
of a simple genetic network in a cell population using rate equations.
In Sec. 3 we consider the limit in which each cell contains a small
population of proteins, where the stochastic features become significant.
The master equation for this system is presented.
The analogy between genetic networks and grain-surface chemistry is
discussed in Sec. 4.
A summary is presented in Sec. 5.

\section{Rate equations}
\label{Sec:rate}

In genetic autoregulatory circuits the production rate 
of a certain product protein $A$ depends on its population size,
$[A]$ (given by the {\it average} number of such proteins in a cell). 
In negative autoregulation, increasing the population size
$[A]$ decreases the rate of production. 
This mechanism is commonly approximated 
\citep{Rosenfeld2002,Paulsson2002} by the Hill function

\begin{equation} 
g(A)={g_{\rm max} \over {1+k[A]}}
\end{equation}

\noindent
where $g(A)$ is the production rate of $A$ proteins,
$g_{\rm max}$ is the maximal production (achieved 
in conditions where $[A]=0$) and $k$
is an affinity constant. 
This approximation is in agreement with experiments
done at steady state 
\citep{Yagil1971,Yagil1975}. 
Here we consider the following circuit:
a population size $[R]$ of mRNA's is produced 
with a maximal rate $g_R$ and degrades at rate 
$d_R$. 
This mRNA produces a protein $A$ which acts as a repressor 
and controls the production rate of the mRNA. 
The production rate of $A$ is thus proportional 
to $[R]$ and its degradation rate is $d_A$.
The intracell dynamics is described by the
rate equations 

\begin{eqnarray}
\label{Eq:rate_2}
\dot{[R]}&=&{g_{R}\over 1+k[A]}-d_R[R]\nonumber\\
\dot{[A]}&=&g_A[R]-d_A[A].
\end{eqnarray}

\noindent
where the dots represent time derivatives, namely
$\dot{[R]} = d[R]/d t$.
These equations have two steady state solutions, however,
only one of them is relevant because the other exhibits
negative population sizes.
The relevant solution is

\begin{eqnarray}
\label{Eq:rate_2_ss}
{[R]}&=&{d_A\over 2g_A}\left[\sqrt{{1 \over k^2}+
{4g_Ag_R \over d_Ad_Rk}}-{1\over k}\right]\nonumber\\
{[A]}&=&\half\left[\sqrt{{1 \over k^2}+
{4g_Ag_R \over d_Ad_Rk}}-{1\over k}\right]
\end{eqnarray}

\noindent
and the convergence to this solution is fast 
\citep{Rosenfeld2002}.
However, these equations do not take into account explicitly
the chemical mechanism
which enables the regulation. 
In this mechanism, one
of the $A$ proteins bounds to the repression site on the 
DNA and inhibits the mRNA production.
This protein should be subtracted from the population of 
free proteins in the cell, 
which Eq.~(\ref{Eq:rate_2}) does not do.
In addition,
the constant $k$ in the
Hill function captures only the steady state repression rate
and not its dynamical behavior.

The dynamics of the repression mechanism can be incorporated into 
the rate equation by taking the bound protein as a 
third component in the reaction network. 
This gives rise to three dynamic equations:

\begin{eqnarray}
\label{Eq:rate_3}
\dot{[R]}&=&g_{R}(1-[r])-d_R[R]\nonumber\\
\dot{[A]}&=&g_A[R]-d_A[A]-\alpha_0[A](1-[r]) + \alpha_1 [r]\\
\dot{[r]}&=&\alpha_0[A](1-[r])-\alpha_1 [r] \nonumber
\end{eqnarray}

\noindent
where $[r]$ represents the average population of bound repressors
in a cell.
Since there is only a single repression site
in each cell, $[r]$ is limited to the
range 
$0 \le [r] \le 1$.
In fact, it represents the fraction of time in 
which the repressor site on the DNA
is occupied by a bound repressor. 
The average productivity of the DNA in producing mRNA's is proportional
to $1-[r]$. 
The binding coefficient
$\alpha_0$ is the rate in which a free protein
becomes bound.
This rate should be 
multiplied by the number of free proteins  
and by the average number of unoccupied repression sites per cell, 
$1-[r]$.
The desorption coefficient 
$\alpha_1$ is the rate in which a 
bound protein leaves the repression site.
The reduced rate equation set given by Eq.  
(\ref{Eq:rate_2}) 
is an approximation to the extended set of
Eq.
(\ref{Eq:rate_3}) 
in the following manner: 
when $\alpha_0$ and $\alpha_1$ are large 
compared to other rate constants, 
$[r]$ approaches steady state much faster than  
$[A]$ and $[R]$. 
In this case, it is justified to assume that $[r]$  
is in quasi steady state
and impose $\dot{[r]}=0$. 
This gives the steady state solution 

\begin{equation}
[r]={\alpha_0[A] \over \alpha_1+\alpha_0[A]}. 
\end{equation}

\begin{figure}
\hspace{-2.0cm}
\includegraphics[angle=270, width=18cm]{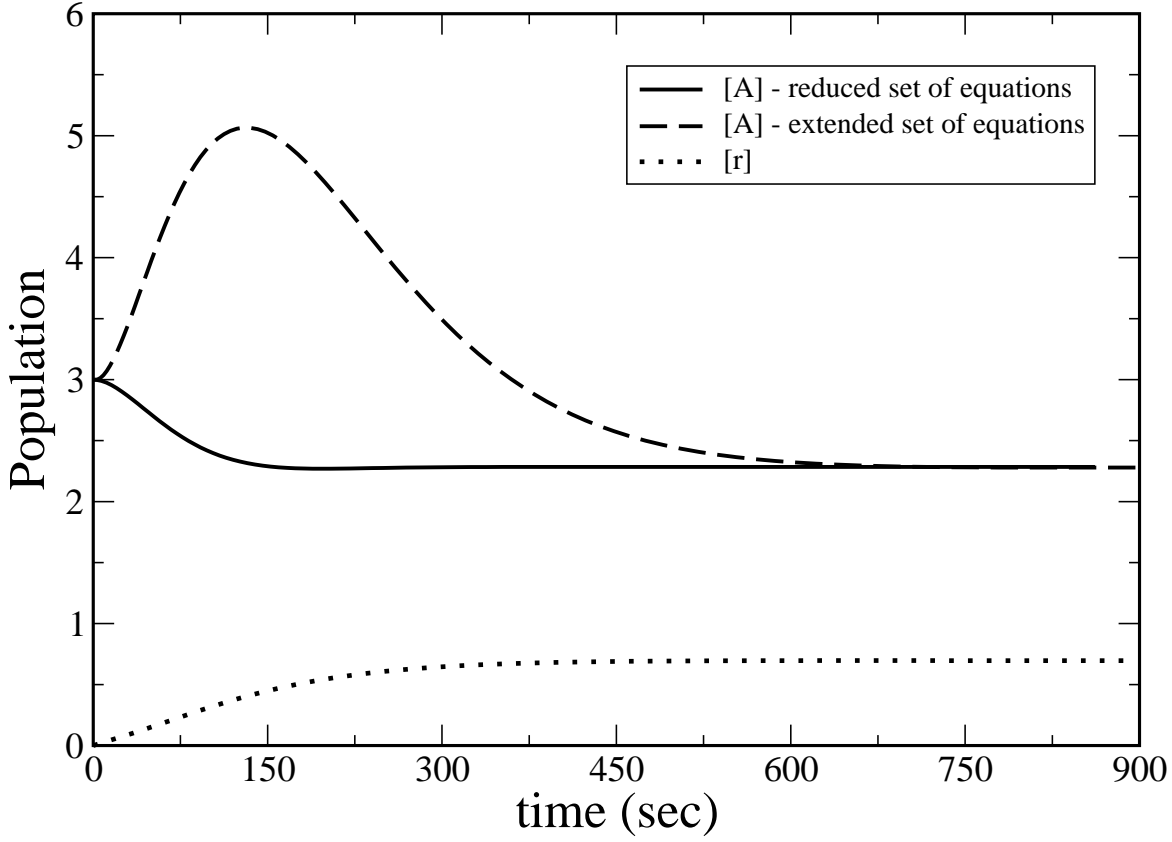}
\caption{Intracell dynamics as calculated by 
Eq. ~(\ref{Eq:rate_2}) (solid line)
and by Eq. ~(\ref{Eq:rate_3}) (dashed line). 
The average amount of bound proteins $[r]$
is also shown (dotted line). 
The initial conditions are $[A]=3$ and $[r]=0$.}
\label{Fig:dynamicsA}
\end{figure}

\begin{figure}
\hspace{-2.0cm}
\includegraphics[angle=270, width=18cm]{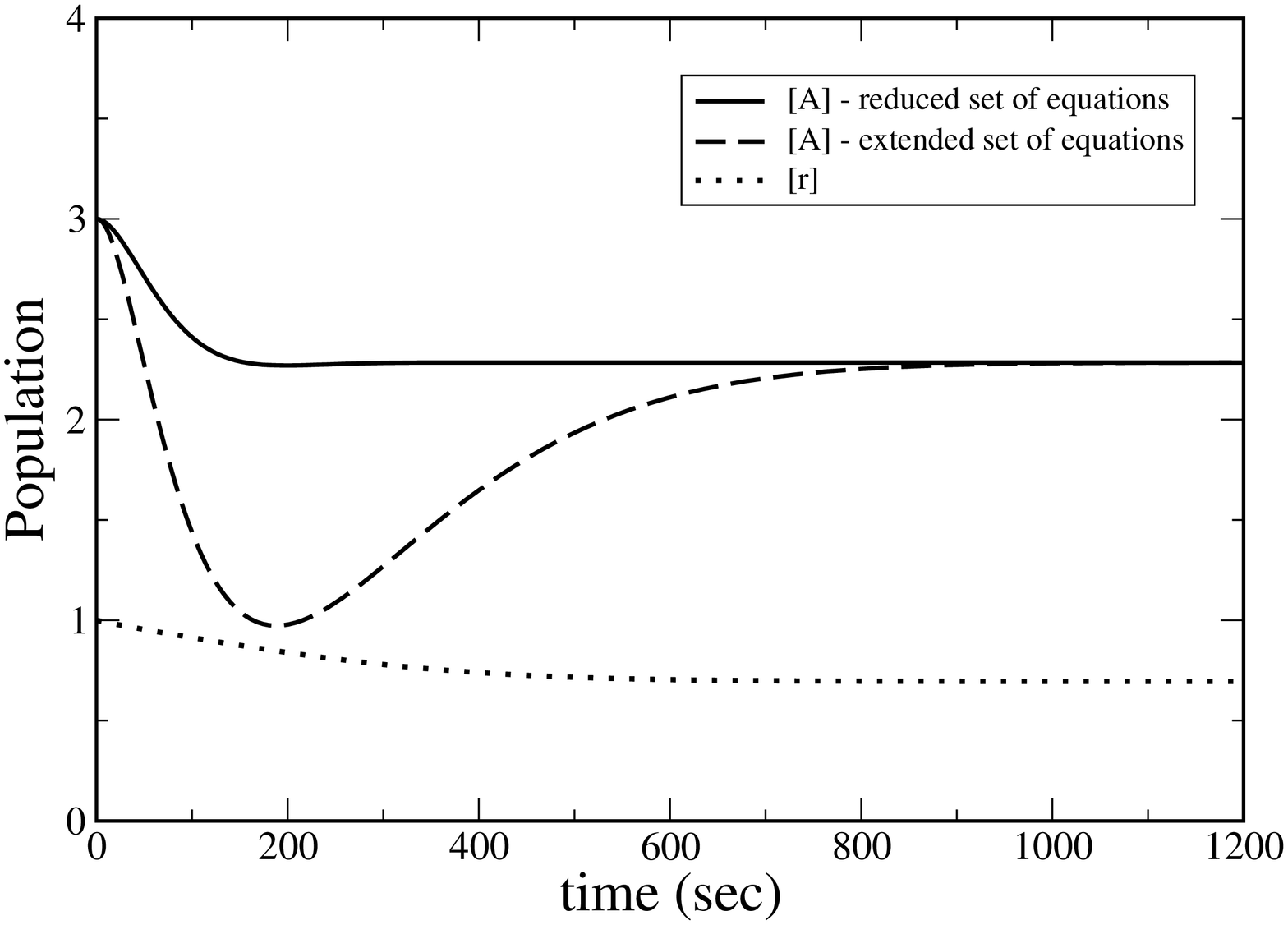}
\caption{Intracell dynamics as calculated by 
Eq. ~(\ref{Eq:rate_2}) (solid line)
and by Eq. ~(\ref{Eq:rate_3}) (dashed line). 
The average amount of bounded proteins $[r]$
is also shown (dotted line). The initial conditions are 
$[A]=3$ and $[r]=1$.}
\label{Fig:dynamicsB}
\end{figure}

\noindent
Substituting this solution into 
Eq. (\ref{Eq:rate_3}) 
gives the reduced set of Eq. (\ref{Eq:rate_2}), 
with $k=\alpha_0/\alpha_1$. 
This implies that Eq. (\ref{Eq:rate_2_ss}) is the  
steady state solution of Eq. (\ref{Eq:rate_3}) as well.
This solution is stable and there are no oscillations
for any values of the parameters. 
However, the time dependent solutions of 
Eq. (\ref{Eq:rate_2}) 
and of 
Eq. (\ref{Eq:rate_3}) 
are not the same.
Whereas Eq. 
(\ref{Eq:rate_2}) 
assumes 
rapid convergence of $[r]$ into its steady state, 
Eq. (\ref{Eq:rate_3}) holds also in case that the relaxation time is long.
In Figs. 
\ref{Fig:dynamicsA} 
and 
\ref{Fig:dynamicsB}   
we compare the dynamics described by the two sets 
of equations. 
The rate constants are 
$g_R = 0.05$, 
$g_A = 0.06$, 
$d_R = 0.02$, 
$d_A = 0.02$, 
$\alpha_0 = 0.001$
and 
$\alpha_1 = 0.001$
(all in units of $s^{-1}$).
These rates represent typical transcription and translation times,
which are of the order of 10 to 20 seconds.
Typical half-life times of proteins and mRNA's vary in the range
of several minutes
\citep{Elowitz2000}.
All these time scales are much shorter than the cycle time, which
is typically around 30 minutes.

The dynamical behavior of $[A]$ turns out to be different in the two
sets of equations. 
The deviations from steady state are 
much larger in the extended set of equations. 
The dynamics is also highly dependent 
on the initial condition of $[r]$ which is an additional 
degree of freedom that does not exist in the reduced set.
In 
Fig.~\ref{Fig:dynamicsA} 
where the initial condition is $[r]=0$,
the extended set shows an over-shoot in $A$ production,
while 
in 
Fig.~\ref{Fig:dynamicsB}  
where the initial condition is $[r]=1$,
it shows an under-shoot in $A$ production.

In some cases the regulation of the production of a protein 
$A$ is mediated by a more complex molecule.
For example, the repressor may be a molecule
D which is a dimer of $A$ molecules
produced by the reaction $A+A \rightarrow D$.
The standard way of modeling such a 
circuit is to modify the repression term 
(the Hill function) in 
Eq. ~(\ref{Eq:rate_2}) to 

\begin{eqnarray}
\label{Eq:rate_2_n}
\dot{[R]}&=&{g_{R} \over 1+k[A]^2}-d_R[R] \nonumber \\
\dot{[A]}&=&g_A[R]-d_A[A].
\end{eqnarray}

\noindent
For this system
the extended set includes equations for
$[R]$ and $[A]$, 
as well as for the
dimer (repressor) population
$[D]$ and for 
the bound repressor $[r]$.
The equations take the form:

\begin{subeqnarray}
\label{Eq:rate_4_n}
\dot{[R]}&=&g_{R}(1-[r])-d_R[R]\\
\slabel{Eq:rate_4_n_R}
\dot{[A]}&=&g_A[R]-d_A[A] - 2 \alpha_2 [A]^2\\
\slabel{Eq:rate_4_n_A}
\dot{[D]}&=& \alpha_2 [A]^2 - d_D [D]-\alpha_0[D] (1-[r])+\alpha_1 [r]\\
\slabel{Eq:rate_4_n_D}
\dot{[r]}&=&\alpha_0[D](1-[r])-\alpha_1 [r]
\slabel{Eq:rate_4_n_r}
\end{subeqnarray}
    
\begin{figure}
\hspace{-2.0cm}
\includegraphics[angle=270, width=18cm]{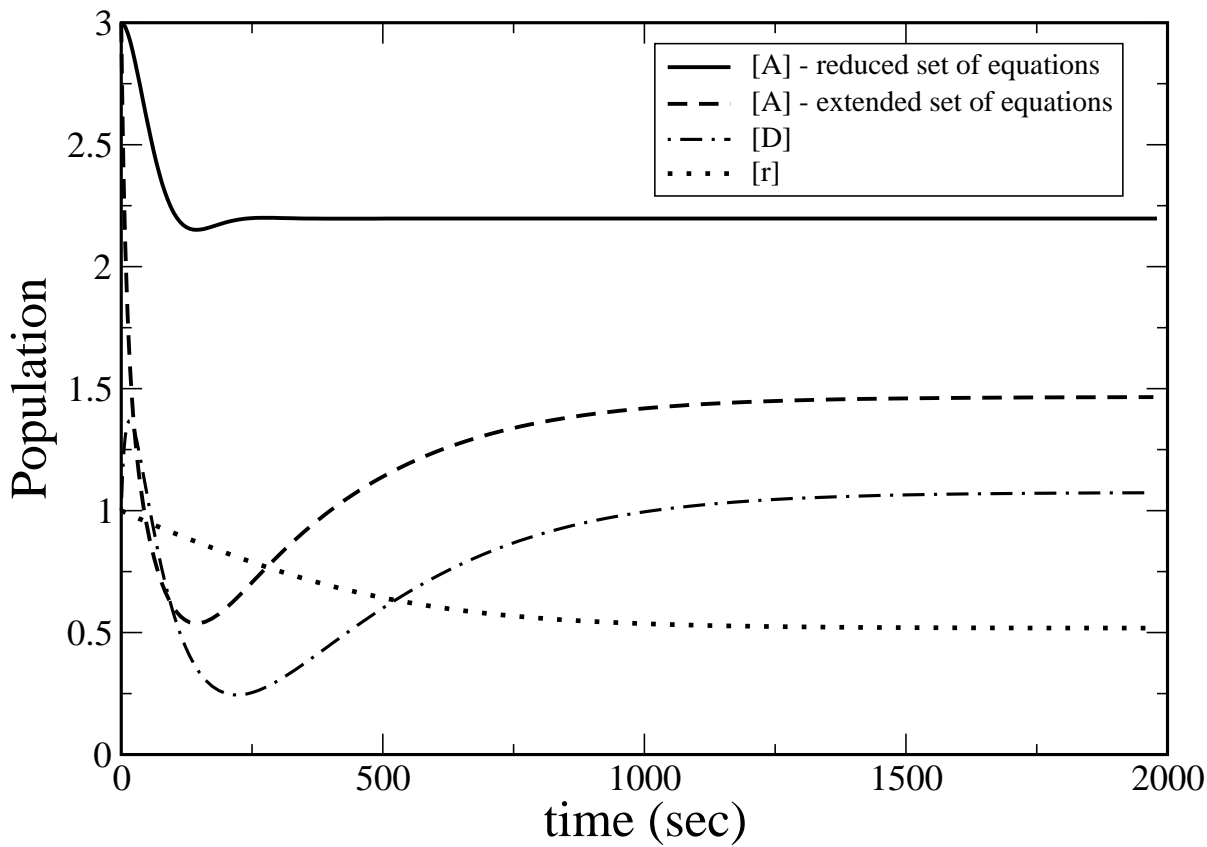}
\caption{The populations of proteins $A$ as obtained from the
reduced set (solid line) and from the extended set (dashed line)
and of dimers (repressor) $D$ (dashed-dotted line) and bound repressors
$r$ (dotted line) as obtained from the extended set, as a function of time.
The initial conditions are $[A]=3$, 
$[D]=1$ and $[r]=1$.
}
\label{Fig:dynamics_n}
\end{figure}

\noindent
As in the ordinary case in which the repressor is the protein $A$ itself,
the inhibition term $1-[r]$ 
in Eq. 
(\ref{Eq:rate_4_n_R}) 
is equal to
the Hill function of Eq.  
(\ref{Eq:rate_2_n})  
in the limit of rapid relaxation of $[r]$.
In this case
$k = \alpha_0 \alpha_2 / (\alpha_1 d_D)$.
However, 
when the repressor is the dimer $D$, there is an additional 
term in Eq. ~(\ref{Eq:rate_4_n_A})
which has no analogue in  Eq. ~(\ref{Eq:rate_2_n}). 
This term gives rise to a difference 
in the results of the reduced and the extended
sets even
in the steady state solution, 
as shown in 
Fig~\ref{Fig:dynamics_n}. 
The steady state solution of the extended set is stable
and exhibits no oscillations.
The parameters used in
Fig~\ref{Fig:dynamics_n} 
are the same as in 
Figs.~\ref{Fig:dynamicsA} 
and
\ref{Fig:dynamicsB}, 
and the additional parameters 
are 
the degradation rate of dimers,
$d_D=0.02$
($s^{-1}$),
and the production rate coefficient of dimers,
$\alpha_2=0.01$
($s^{-1}$).
The latter coefficient is determined by the diffusion
rate of proteins in the cell.
A related quantity, namely, the time it takes for a protein
to diffuse across the cell was recently measured
\citep{Elowitz1999}
and found to be of the order of one second.
The inverse of this time can be used as an upper 
bound for the production rate coefficient $\alpha_2$.

The reduced set of equations does not 
take into account explicitly the dimer population,
which is responsible for the repression.
Both Eqs.
(\ref{Eq:rate_2_n})
and 
(\ref{Eq:rate_4_n})
do not take into account the fact that one needs
at least two $A$ proteins 
simultaneously in the cell
in order to produce a dimer.
Therefore, when the population of $A$ proteins goes down
to order 1 both equations fail
and the master equation formalism is required.

\section{The Master Equation}
\label{Sec:master}

Rate equations are used to describe the dynamics of the average number of 
entities (such as proteins)
in large populations such as those handled in {\it in vitro} experiments. 
In these equations it is assumed that the densities of substances 
are continuous variables that behave in a deterministic fashion.
This approach is 
not suitable for genetic regulatory
networks when the populations of the relevant species in a single
cell are small
\citep{Gillespie1977,Nicolis1977,Ko1991,Ko1992,McAdams1999,Szallasi1999,Gibson2001}.
In this case one should take into account the discrete nature of the
populations and the fact that for small populations the fluctuations
become significant.
In negative regulatory systems there is a population of free repressors
in the cell. In addition, there is a single repression site
on the DNA where a single repressor molecule may bound. 
Therefore, each repression site can be either 
occupied by a repressor molecule (where $r=1$)
or vacant ($r=0$).
Thus, $r$ cannot take any intermediate values.
In such cases fluctuations may have 
an important impact on the processes involved  and their dynamics 
should be described in more detail. 

One of the approaches suggested is the 
use of stochastic simulations
which take into account the dynamics of all
participating substances
\citep{Gillespie1977,McAdams1997,Morton1998,Gibson2000}. 
The difficulty with these simulations is that they 
are based on the accumulation of large amounts of
statistical data, and thus require extensive computer simulations.  
Thus, this approach is 
not always feasible in the case of complex networks 
which involve a large number of proteins. 
A complementary approach is based on 
direct integration of the the master equation
\citep{McAdams1997,McAdams1999,Paulsson2000,Paulsson2000b,Kepler2001,Paulsson2002,Paulsson2004}. 
This approach takes into account the probability distribution 
of all possible states of the system, and not only 
the average values as in the rate equation approach.
It captures the time evolution of the probabilities
of all the microscopic states of the system.

We now apply the master equation approach to study the 
negative autoregulatory circuit of Eq. 4.
We denote the number of copies of the free protein $A$ by
$n_A$ and of the mRNA by $n_R$.
The number of proteins $A$ which are bound to the repression site on the
DNA is given by $n_r$.
For a single repression site $n_r$ can only take the values 0 or 1. 
The master equation follows the time evolution of the probability 
distribution
$P(n_R,n_A,n_r)$.  
It takes the form

\begin{subeqnarray}
\dot P(n_R,n_A,n_r=1) &=& 
g_A n_R [P(n_R,n_A-1,1)-P(n_R,n_A,1)] \nonumber \\
&+& d_R[(n_R+1)P(n_R+1,n_A,1)-n_RP(n_R,n_A,1)] \nonumber \\
&+& d_A[(n_A+1)P(n_R,n_A+1,1)-n_AP(n_R,n_A,1)] \nonumber \\
&+& \alpha_0((n_A+1)P(n_R,n_A+1,0) \nonumber \\
&-& \alpha_1P(n_R,n_A,1)\\
\slabel{Eq:master1}
\dot P(n_R,n_A,n_r=0) &=&  
g_A n_R [P(n_R,n_A-1,0)-P(n_R,n_A,0)] \nonumber \\
&+& d_R[(n_R+1)P(n_R+1,n_A,0)-n_R P(n_R,n_A,0)]\nonumber \\
&+& d_A[(n_A+1)P(n_R,n_A+1,0)-n_A P(n_R,n_A,0)]\nonumber \\
&-& \alpha_0 n_A P(n_R,n_A,0)\nonumber \\
&+& \alpha_1 P(n_R,n_A-1,1) \nonumber \\
&+& g_R[P(n_R-1,n_A,0)-P(n_R,n_A,0)], 
\slabel{Eq:master2}
\end{subeqnarray}

\noindent
where the two cases of $n_r=0$ and $n_r=1$ are
presented separately. 
The first terms in the equations describe
the formation of a new protein. 
The second and third terms describe the
degradation of the mRNA and the protein, respectively, while the fourth
and fifth terms describe the binding and unbinding of a protein to the
repression site on the DNA. 
Eq. ~(\ref{Eq:master2}) 
also includes
a term that corresponds to the formation of a new mRNA 
(not possible
in the repressed case). 
These equations can be integrated numerically 
in order to obtain the time dependence of
the probability distribution.
It can also be
solved for steady state by taking 
$\dot P(n_R,n_A,n_r) = 0$. 

The master equation provides all the moments of the
distribution
$P(n_R,n_A,n_r)$
and their time dependence. 
For example,
the average population of proteins $A$
is given by

\begin{equation}
\langle{n_A}\rangle=
\sum_{n_R=0}^{n_R^{\rm max}} 
\sum_{n_A=0}^{n_A^{\rm max}}  
\sum_{n_r=0}^{1} 
n_A P(n_R,n_A,n_r)
\label{eq:average}
\end{equation}

\noindent
where 
$n_R^{\rm max}$  
and 
$n_A^{\rm max}$  
are the cutoff values that provide upper bounds on the populations of mRNA 
molecules and $A$ proteins in the cell, respectively. 
The repression site can
be either occupied ($n_r=1$) or unoccupied ($n_r=0$). 

\begin{figure}
\center
\hspace{-2.0cm}
\includegraphics[width=12cm]{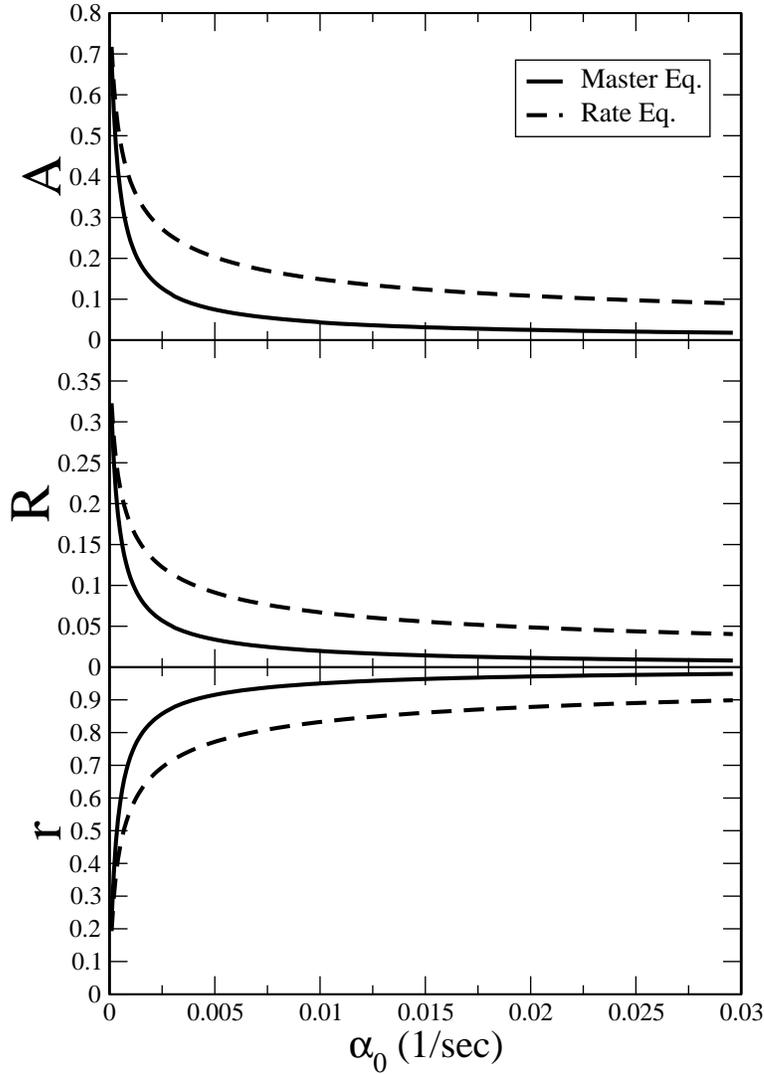}
\caption{The steady state populations of free proteins, 
mRNA's and bound proteins (repressor) vs. the rate constant
$\alpha_0$, calculated using the master 
equation (solid line) and the rate equations (dashed line).}
\label{Fig:comparison}
\end{figure}

\noindent
Solving the master equation under steady state conditions
for systems with different rate
constants we calculated the appropriate averages, and compared
the results with the 
rate equations.

In Fig. \ref{Fig:comparison} 
the average levels of free proteins, 
mRNA molecules and bound protein (repressor) in the cell (at steady state),
are shown vs. $\alpha_0$, as obtained from the  
master equation (solid line) and the rate equations 
(dashed line).
The rate equations turn out to overestimate the average 
level of proteins and mRNA 
molecules, by a factor of 2-4 for systems with low copy number
of proteins. 
On the other hand, when the average number of proteins 
in the cell is large, the results of the rate equations and master equation
coincide. 

Mathematically the discrepency between the results of the
rate equations and the master equations is due to non-linear
terms
such as the term that describe the attachment rate of proteins
to the repression site.
In the rate equation, this term is given by
$\alpha_0 [A] (1-[r])$,
namely as a product of averages (first moments). 
In the master equation it is given by
the second moment
$\alpha_0 \langle n_A (1-n_R) \rangle$.
The formation of dimers is also described by a 
nonlinear term.
In the rate equations this term is given by
$\alpha_2 [A]^2$, namely it depends only on the first moment.
In the master equation it
is given by
$\alpha_2 \langle n_A^2 \rangle - \alpha_2 \langle n_A \rangle$,
thus it depends on both the first and second moments.

The simple networks studied here can be considered as modules
or motifs in complex genetic networks. However, the simulation
of complex networks using the master equation is difficult.
This is due to the proliferation in the number of equations
as the number of components (mRNA's and proteins) increases.
Consider, for example, a network that involves three protein
species, A, B and C. 
The master equation is written in terms
of the probabilities
$P(n_A,n_B,n_C)$ 
of having a certain population of proteins.
The population size of each protein is limited
by an upper cutoff. For example, the population
of protein A takes the values $n_A = 0,1,\dots,n_A^{\rm max}$.
Clearly, the number of equations increases exponentially 
with the number of species, making this approach infeasible
for complex networks.
 
However, typically these networks are sparse, namely most 
pairs of proteins do not interact with each other. 
This feature makes it possible to divide the master equation
into several sets of equations, each set including only a
small number of protein species. 
For example, if proteins B and C do not interact, the master
equation described above can be broken into two sets that
involve 
$P_{\rm AB}(n_A,n_B)$
and 
$P_{\rm AC}(n_A,n_C)$.
In the case of large and sparse networks this dramatically
reduces the number of equations and thus enables the simulation
of complex networks using the master equation.
This technique, named the multi-plane method, was recently proposed
in the context of chemical reaction networks on interstellar
dust grains
\citep{Lipshtat2004}. 
The mathematical structure of these networks is similar
to that of genetic networks. Thus, the multi-plane method is
perfectly applicable for the simulations of complex genetic networks.
The similarity between the two systems is briefly discussed below.

\section{Discussion: Genetic Networks and Grain-Surface Chemistry}

Processes which exhibit a similar mathematical structure
to the genetic network dynamics appear in the context
of chemical reaction networks on interstellar dust grains.
The chemistry of interstellar clouds consists of reactions taking
place in the gas phase as well as on the surfaces of dust grains
\citep{Hartquist1995}.
It turns out that the most abundant molecule in the Universe,
namely molecular hydrogen does not form in the gas phase but
on dust grain surfaces
\citep{Gould1963,Hollenbach1971a,Hollenbach1971b}. 
These grains are made of amorphous silicate and carbon 
compounds and are of sub-micron size. 
In addition to the formation of molecular hydrogen, these grains
support complex reaction networks that produce a variety of molecules
that consist of hydrogen, oxygen, carbon and nitrogen.
Here we discuss the similarity between 
the mathematical descriptions surface reaction networks and genetic networks.
In particular, we suggest that computational methodologies developed 
in the context of interstellar grain chemistry are likely to be 
useful for the analysis of genetic networks.

\begin{table}
\caption{Analogy between the processes of surface chemistry and
of gene regulation. }
\begin{tabular}{lll}
Description & Surface chemistry & Gene regulation\\
\hline
system &  dust grain & cell \\
break-up mechanism & grain fragmentation & cell division \\
mobility & surface diffusion & diffusion in cell \\
addition $\varnothing\rightarrow A$ & flux $F$ & transcription $g_R$, 
production $g_A$\\
removal $A \rightarrow \varnothing$ & desorption $W$ & degradation 
$d_R,d_A$ \\
typical reaction & $A+B\rightarrow C+D$  & $A\rightarrow A+B$\\
feedback regulation & rejection: $F(1 - \theta)$ & repression: $g_R(1-[r])$ \\
\hline

\end{tabular}
\end{table}

Consider a dust grain 
exposed to a flux of atomic and molecular species such
as H, O, OH and CO.
Atoms and molecules that hit and stick to the grain 
hop as random walkers between adsorption sites 
on its surface.
When two atoms/molecules encounter one another they
may react and form a more complex molecule.
The rate equations that describe the reaction networks
on grains include flux terms, desorption terms and reaction 
terms.
The flux terms represent the flow of atoms and molecules 
from the gas phase onto the surface. 
The desorption rates are proportional to the population
sizes of atoms and molecules on the grains, while the
reaction terms are proportional to the products of the population
sizes of the reactive species.  
In general, the rate equations resemble those that describe 
genetic networks.
The analogy between the two systems is summarized in Table 1.
In both systems reactive species are added, diffuse, react 
and removed.
The system itself may break up (cell division or grain fragmentation),
dividing the population of reactive species into two sub-populations.
Both systems exhibit some kind of negative feedback.
In genetic networks this is provided by the repression circuit,
in which the rate of attachment of proteins to the repression
site is given by
$\alpha_0 [A] (1-[r])$.
Certain surface reaction systems exhibit the Langmuir rejection
behavior, in which atoms from the gas phase that hit the surface
in the vicinity of an already adsorbed atoms are rejected.
The flux term $F$ is then modified to the form
$F(1 - \theta)$, where $\theta$ is the coverage, namely the
fraction of adsorption sites on the surface that are occupied
by adsorbed atoms.
In the context of grain-surface chemistry, low
copy numbers are obtained in the limit of small grains 
under conditions of low flux. 
In this limit the master equation is required
\citep{biham2001,Green2001,Biham2002}.
For complex reaction networks of multiple species, 
the master equation becomes infeasible due to the
proliferation in the number of equations. 
In this case, the multi-plane method is used in order
to keep the number of equations at a tractable level
\citep{Lipshtat2004}.

\section{Summary}

We have considered the 
rate equation and master equation approaches to the 
modeling of genetic networks. 
In particular, we have studied
the temporal evolution of the population of mRNA and proteins
in simple negative autoregulated genetic networks. 
As long as the populations of all the reactive components
of the network are not too small, rate
equations provide a good quantitative description of the
network dynamics.
However, once the populations of the mRNA or proteins
are reduced to order 1 or less, 
rate equations are no longer suitable and the master equation
is needed.
This is due to the fact that the rate equations involve
only average quantities, while the master equation takes into
account the discrete nature of the populations as well as
the fluctuations.
The simple networks studied here can be considered as
modules or motifs in complex genetic networks.
The simulation of complex networks using the master equation
is difficult, because the number of equations quickly proliferates.
The multi-plane methodology, recently developed in the
context of grain-surface chemistry,
that tackles this problem is briefly described.
Finally,
the analogy between genetic
networks and grain-surface chemistry is discussed. 

We thank J. Paulsson for illuminating discussions.









\end{document}